\documentclass[11pt,a4paper]{article}
\pdfoutput=1 
\usepackage{jheppub}	
\usepackage{tikz}
\usetikzlibrary{patterns}
\usetikzlibrary{positioning,arrows}
\usetikzlibrary{decorations.pathmorphing}
\usetikzlibrary{decorations.markings}
\usetikzlibrary{snakes}
\usetikzlibrary{matrix}

\usepackage{amsmath}
\usepackage{bm}
\usepackage[utf8]{inputenc}
\usepackage[english]{babel}
\usepackage{makeidx}
\usepackage{tikz}
\tikzset{every picture/.style={line width=0.75pt}} 
\usepackage{amsfonts}
\usepackage{enumerate}
\usepackage{mathrsfs}
\usepackage{tensor}
\usepackage[autostyle]{csquotes}
\usepackage{subfig}

\def\be{\begin{equation}}
	\def\ee{\end{equation}}
\def\bea{\begin{eqnarray}}
	\def\eea{\end{eqnarray}}
\newcommand{\nn}{\nonumber}

\def\apjl{\ref@jnl{ApJ}}

\newcommand{\RNum}[1]{\uppercase\expandafter{\romannumeral #1\relax}}

\definecolor{grayfill}{RGB}{230,230,230}
\definecolor{grayline}{RGB}{140,140,140}
\definecolor{mauvefill}{RGB}{226,214,223}
\definecolor{mauveline}{RGB}{128,92,116}

\usepackage{amsmath,amssymb}
\usepackage{latexsym}
\usepackage{graphicx}

\usepackage{slashed}

\usepackage[vcentermath,enableskew]{youngtab}
\usepackage{epsfig}

\def\be{\begin{equation}}
	\def\ee{\end{equation}}
\def\bea{\begin{eqnarray}}
	\def\eea{\end{eqnarray}}

\title{The $W_n$ Light One-Point Torus Conformal Block}

\author[]{Armen Poghosyan and}
\author[]{Hasmik Poghosyan}
\emailAdd{armenpoghos@gmail.com}
\emailAdd{h.poghosyan@yerphi.am}
\affiliation[]{Yerevan Physics Institute \\
	Alikhanian Br. 2, 0036 Yerevan, Armenia}

\abstract{
	We study the light asymptotic limit of the one-point torus conformal block in $A_{n-1}$ Toda field theory. Through the AGT correspondence, this problem can be translated into the computation of the instanton partition function of four-dimensional ${\cal N}=2^{\ast}$ $U(n)$ supersymmetric Yang--Mills theory, which we then examine in the limit $b\to 0$ at fixed conformal dimensions. We show that, in this regime, the instanton sum simplifies drastically: for each Young diagram, only boxes with specific arm lengths contribute to the bifundamental factors. Exploiting this property, we derive an explicit representation for the light one-point torus $W_n$ conformal block valid for arbitrary $n\ge 2$. 
	
	As a consistency check, we specialize our construction to the Liouville case $n=2$ and compare it with the previously known hypergeometric representation of the torus block in the light limit. We also discuss the $W_3$ case and its relation to a known alternative representation obtained by the shadow formalism. 
}
 \clearpage
\begin{document}
	\tikzset{
		line/.style={thick, decorate, draw=black,}
	}
	
	\maketitle
	

\section{Introduction}
Let us consider $A_{n-1}$ Toda field theories \cite{Zamolodchikov:1985wn}, 
which provide a higher-rank generalization of Liouville field theory. 
Recall that Liouville theory \cite{Polyakov:1981xx} is a bosonic two-dimensional conformal field theory \cite{BELAVIN1984333} 
characterized by an exponential potential. 
Its conformal invariance implies the existence of conserved spin-two currents, 
namely the holomorphic and anti-holomorphic components of the stress-energy tensor. 
In the Toda case, the symmetry structure is richer: in addition to the spin-two currents, 
the theory also admits conserved currents of higher spin. 
The corresponding extended symmetry algebra is known as the ${\cal W}$-algebra.

The main object of interest in this paper is the light conformal block, which arises in the classical, 
or equivalently large-central-charge, limit of Toda theory, in which the conformal dimensions of the primary fields remain finite.

To derive the one-point $W_n$ torus conformal block in the light limit, we use the well-known AGT duality \cite{Alday:2009aq,Alba:2010qc}, 
which relates Liouville conformal blocks to the instanton partition functions \cite{Losev:1997bz,Nekrasov:2002qd,Nekrasov:2003rj} 
of four-dimensional ${\cal N}=2$ supersymmetric gauge theories. 
This correspondence was later generalized to $A_{n-1}$ Toda field theories in \cite{Wyllard:2009hg}.

Thanks to this duality, conformal blocks can be extracted from the study of instanton partition functions. 
In particular, an explicit expression for the four-point $W_n$ conformal block in the light asymptotic limit was obtained in \cite{Poghosyan:2016lya} 
using instanton-counting methods \cite{Flume:2002az,Bruzzo:2002xf}. 
It was shown that, in this limit, the instanton partition function simplifies drastically, 
leading to a closed-form expression for the corresponding conformal block.

While the results of \cite{Poghosyan:2016lya} concern the four-point conformal block on the sphere, 
the present work is devoted to the one-point conformal block on the torus. 
The key observation underlying our result is that, in the light asymptotic limit, 
the instanton partition function of the ${\cal N}=2^\ast$ theory also simplifies considerably. 
This simplification makes it possible to obtain a closed-form expression for the partition function and, 
via the AGT correspondence \cite{He:2012bi,Poghossian:2017atl}, to derive the one-point torus conformal block in the light limit.

The fact that our formula is valid for generic $n$ allows us to study its behavior at large $n$. 
This limit is of particular interest in the context of $AdS_3 / CFT_2$ holography \cite{Henneaux:2010xg,Gaberdiel:2010ar,Gaiotto:2026qai}.

The paper is organized as follows. In Section~\ref{revue}, we review \(W_n\) conformal field theory, instanton counting, and the AGT correspondence. In Section~\ref{Zlight}, we show that, in the light limit, only certain boxes contribute to the product in (\ref{Fy}) for a given set of Young diagrams \(\vec{Y}\). As a consequence, the instanton partition function simplifies significantly. In Section~\ref{Lblock}, we present a detailed derivation of the Liouville conformal block using our approach and compare our result with the expression previously obtained in the literature~\cite{Hadasz:2009db,Alkalaev:2016fok}. Finally, in Section~\ref{Wblock}, we present our final formula for the \(W_n\) one-point light block. In Appendix~\ref{W3block}, we provide the \(W_3\) one-point light block in the representation obtained in \cite{Belavin:2024nnw}.
\section{Preliminaries on $A_{n-1}$ Toda Theory, ${\cal N}=2^{\ast}$ SYM Instanton Counting and AGT Duality}	
\label{revue}
\subsection{$W_n$ conformal field theories}
\label{WnTheory}
These are two-dimensional CFTs which, in addition to the spin-2 holomorphic
energy-momentum tensor $W^{(2)}(z)\equiv T(z)$, possess additional
higher-spin currents
$s=3,4,\ldots,n$, namely $W^{(3)}(z), \ldots, W^{(n)}(z)$, with Virasoro
central charge conventionally parametrized as
\[
c=n-1+12 \bm{Q}^2\, ,
\]
where the background-charge vector is
\[
\bm{Q}=Q \bm{\rho},
\qquad
Q=b+\frac{1}{b},
\]
with $\bm{\rho}$ the Weyl vector of the algebra $A_{n-1}$ and $b$ the
dimensionless coupling constant of Toda theory. In what follows, it will
be convenient to represent the roots, weights, and Cartan elements of $A_{n-1}$
as $n$-component vectors with the usual Kronecker scalar product, subject
to the condition that the sum of their components vanishes. This is equivalent
to the more conventional representation of these quantities as diagonal traceless
$n\times n$ matrices with the pairing given by the trace. In this representation,
the Weyl vector is given by
\bea\label{weylvec}
\bm{\rho}=\left(\frac{n-1}{2},\frac{n-3}{2},\ldots,\frac{1-n}{2}\right)
\quad {\rm or} \quad
\rho_u=\frac{n+1}{2}-u
\eea
and the central charge becomes
\[
c=(n-1)(1+n(n+1)Q^2)\, .
\]

For later reference, let us quote the explicit expressions for the highest weight
$\bm{\omega}_1$ of the first fundamental representation and for its complete set of weights
$\bm{h}_1, \ldots, \bm{h}_n$ (with $\bm{h}_1=\bm{\omega}_1$):
\bea
&&(\bm{\omega}_1)_k=\delta_{1,k}-\frac{1}{n}\,, \nonumber\\
&&(\bm{h}_l)_k=\delta_{l,k}-\frac{1}{n}\, .
\eea
The primary fields $V_{\bm{\alpha}}$ (here we restrict ourselves to the left-moving
holomorphic sector) are parametrized by vectors $\bm{\alpha}$ with
vanishing center of mass. Their conformal weights are given by
\bea
h_{\bm{\alpha}}=\frac{\bm{\alpha}(2 \bm{Q}-\bm{\alpha})}{2}\, .
\label{dim_gen}
\eea
In what follows, a special role is played by the fields $V_{\lambda \bm{\omega}_1}$
with conformal dimensions
\bea
h_{\lambda \bm{\omega}_1}=\frac{\lambda(n-1)}{2} \left(Q-\frac{\lambda}{n}\right)\, .
\label{dim_spec}
\eea
The one-point conformal block on the torus with modular parameter $\tau$ is defined as
\bea
\label{1pBlock}
{\cal F}_{\bm{\alpha}}^\lambda (q)=q^{\frac{c}{24}-h_{\bm{\alpha}}} {\rm tr}_{\bm{\alpha}}
\left(q^{L_0-\frac{c}{24}} V_{\lambda \bm{\omega}_1}(1)\right)\,,
\eea
where  
\bea
\label{q}
q=e^{2 \pi i \tau}\, .
\eea

\subsection{Instanton counting for ${\cal N}=2^{\ast}$ SYM theory}
\label{sec:npfof}

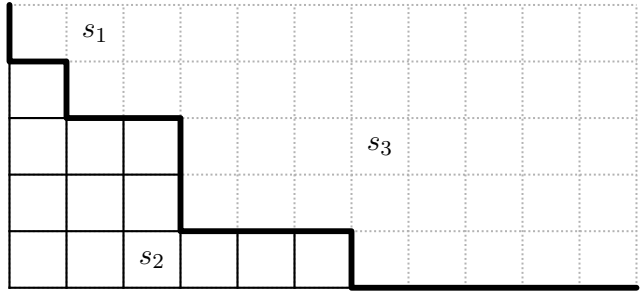
\begin{figure}
	\centering
	\begin{tikzpicture}[
		x=0.75cm,y=0.75cm,
		inner/.style={line width=0.8pt,line cap=rect},
		outer/.style={line width=2.2pt,line join=round,line cap=round}
		]
		\draw[gray!60,densely dotted,step=1] (0,0) grid (11,5);
		
		\draw[inner] (0,0) -- (6,0);
		\draw[inner] (0,1) -- (3,1);
		\draw[inner] (0,2) -- (3,2);
		\draw[inner] (0,3) -- (1,3);
		
		\draw[inner] (0,0) -- (0,4);
		\draw[inner] (1,0) -- (1,3);
		\draw[inner] (2,0) -- (2,3);
		\draw[inner] (3,0) -- (3,1);
		\draw[inner] (4,0) -- (4,1);
		\draw[inner] (5,0) -- (5,1);
		
		\draw[outer]
		(0,5) -- (0,4) -- (1,4) -- (1,3) -- (3,3) -- (3,1)
		-- (6,1) -- (6,0) -- (11,0);
		
		\node[inner sep=0pt] at (1.5,4.5) {$s_1$};
		\node[inner sep=0pt] at (2.5,0.5) {$s_2$};
		\node[inner sep=0pt] at (6.5,2.5) {$s_3$};
	\end{tikzpicture}
	\caption{The arm and leg lengths with respect to the Young diagram (outlined by the thick black line) are
		$A(s_1)=-2$, $L(s_1)=-2$, $A(s_2)=2$, $L(s_2)=3$, $A(s_3)=-3$, and $L(s_3)=-4$.
		This Young diagram is uniquely characterized by the infinite sequence of nonnegative  integers
		$\{3,0,2,1,0,0,0,\ldots\}$.}
	\label{fig:arm_and_leglength}
\end{figure}
Consider ${\cal N}=2^{\ast}$ SYM theory with gauge group $U(n)$ and an adjoint hypermultiplet. The instanton part of the
partition function \cite{Losev:1997bz,Nekrasov:2002qd,Nekrasov:2003rj} of this theory can be written as
\begin{eqnarray}\label{instfu}
	Z_{inst}=\sum_{\vec{Y}}F_{\vec{Y}}q^{|\vec{Y}|},
\end{eqnarray}
where $\vec{Y}$ is an $n$-tuple of Young diagrams,
$|\vec{Y}|$ is the total number of boxes, and $q$ is the instanton-counting parameter related to
the gauge coupling in the standard way. The coefficients $F_{\vec{Y}}$
are given by (see \cite{Flume:2002az,Bruzzo:2002xf})
\begin{eqnarray}
	\label{Fy}
	F_{\vec{Y}}=\prod_{u,v=1}^n
	\frac{
		Z_{bf}(a_u-m,Y_u\mid a_v,Y_v)}
	{Z_{bf}(a_u,Y_u\mid a_v,Y_v)}\,,
\end{eqnarray}
where
\begin{eqnarray}
	\label{Zbf}
	Z_{bf}(a,\lambda\mid b,\mu)&=&\displaystyle\prod_{s\in\lambda}\big(a-b-\epsilon_1L_{\mu}(s)+
	\epsilon_2(1+A_{\lambda}(s))\big)\\
	&\times&
	\displaystyle\prod_{s\in\mu}\big(a-b+\epsilon_1(1+L_{\lambda}(s))
	-\epsilon_2A_{\mu}(s)\big)\,.\nonumber
\end{eqnarray}

Here $A_{\lambda}(s)$ and $L_{\lambda}(s)$ denote, respectively, the arm length and leg length
of the square $s$ with respect to the Young diagram $\lambda$, defined as the oriented horizontal and vertical distances
from the square $s$ to the outer boundary of $\lambda$ (see Fig.~\ref{fig:arm_and_leglength}).

Let us now clarify our conventions for the gauge-theory parameters $a_u$,
	$u=1,2,\ldots,n$. The parameters $a_u$ are expectation values
	of the scalar field in the vector multiplet. Without loss of generality,
	we assume that the ``center of mass'' of these expectation values vanishes:
\bea\label{au1}
\bar a=\frac{1}{n}\sum_{u=1}^n a_u=0\,.
\eea
This is indeed not a restriction, since a nonzero center of mass can be absorbed in the adjoint  hypermultiplet mass $m$.
Finally, $\epsilon_1$ and $\epsilon_2$ are the $\Omega$-background parameters.

Due to AGT duality,
this partition function is directly related to a particular
one-point conformal block in two-dimensional $A_{n-1}$ Toda field theory.
We now describe this relation.
\subsection{AGT duality}

According to AGT duality \cite{Alday:2009aq,He:2012bi,Poghossian:2017atl}, the instanton partition function
$Z_{\mathrm{inst}}$ (see (\ref{instfu}))
is related to the conformal block
${\cal F}_{\bm{\alpha}}^{\lambda}(q)$ (see (\ref{1pBlock})) by
\bea
\label{AGT}
Z_{\mathrm{inst}}
=
\left(q^{-\frac{1}{24}}\eta(q)\right)^{
	\lambda\left(Q-\frac{\lambda}{n}\right)-1}
\,{\cal F}_{\bm{\alpha}}^{\lambda}(q)\, .
\eea
where $\eta (q)$ is Dedekind's eta function
\bea
\eta(q)=q^{\frac{1}{24}}\prod_{n=1}^{\infty}(1-q^n)\,.
\eea
The relation (\ref{AGT}) holds provided that the gauge-theory parameters are identified
with the conformal-block parameters as described below.

First, one identifies the instanton-counting parameter $q$ with the parameter $q$ on the CFT side \eqref{q},
while the Toda coupling $b$ is related to the $\Omega$-background parameters $\epsilon_1$ and $\epsilon_2$ by
\bea
b=\sqrt{\frac{\epsilon_1}{\epsilon_2}}\,.
\label{bepsilon}
\eea
Finally, the correspondence between the gauge-theory parameters $a_u$
and $m$ and the CFT parameters $\bm{\alpha}$ and $\lambda$ is given by
\footnote{Note that the first relation in (\ref{AGTmap}) can be written as
	$\frac{a_u}{\sqrt{\epsilon_1\epsilon_2}}=\left(Q \bm{\rho}-\bm{\alpha}\right)_u$.}
\begin{eqnarray}
	\label{AGTmap}
	\frac{a_u}{\sqrt{\epsilon_1\epsilon_2}}
	&=&
	-\alpha_u
	+
	Q\left(\frac{n+1}{2}-u\right),
	\qquad
	\frac{m}{\sqrt{\epsilon_1\epsilon_2}}
	=
	\frac{\lambda}{n}\, .
\end{eqnarray}

\section{The light asymptotic limit of the instanton partition function}
\label{Zlight}

\begin{figure}
	\centering
	\begin{tikzpicture}[x=9mm,y=9mm,line cap=round,line join=round,scale=0.9]
		
		\tikzset{
			cell/.style={draw=black, thin},
			contour/.style={draw=black, line width=2.4pt}
		}
		
		\newcommand{\emptycell}[2]{%
			\draw[cell] (#1,#2) rectangle ++(1,1);
		}
		\newcommand{\gcell}[2]{%
			\fill[grayfill] (#1,#2) rectangle ++(1,1);
			\fill[pattern=north east lines, pattern color=grayline]
			(#1,#2) rectangle ++(1,1);
			\draw[cell] (#1,#2) rectangle ++(1,1);
		}
		\newcommand{\pcell}[2]{%
			\fill[mauvefill] (#1,#2) rectangle ++(1,1);
			\fill[pattern=north west lines, pattern color=mauveline]
			(#1,#2) rectangle ++(1,1);
			\draw[cell] (#1,#2) rectangle ++(1,1);
		}
		
		\foreach \x/\h in {0/5,1/3,2/3,3/3,4/2,5/1,6/1}{
			\foreach \y in {0,...,\numexpr\h-1\relax}{
				\emptycell{\x}{\y}
			}
		}
		
		\pcell{0}{4}
		\gcell{0}{3}
		
		\pcell{1}{2}
		\pcell{2}{2}
		\pcell{3}{2}
		
		\gcell{1}{1}
		\gcell{2}{1}
		\gcell{3}{1}
		
		\pcell{4}{1}
		\gcell{4}{0}
		
		\pcell{5}{0}
		\pcell{6}{0}
		
		\draw[contour]
		(0,6.15) -- (0,5) -- (1,5) -- (1,3) -- (4,3)
		-- (4,2) -- (5,2) -- (5,1) -- (7,1) -- (7,0) -- (9,0);
		
		\node at (0.55,6.35) {$\vdots$};
		
		\node[right=2pt] at (0.15,5.55) {$\ell_{u,5}=0$};
		\node[right=6pt] at (1.05,4.50) {$\ell_{u,4}=1$};
		\node[right=6pt] at (1.05,3.50) {$\ell_{u,3}=0$};
		
		\node[right=6pt] at (4.05,2.50) {$\ell_{u,2}=3$};
		\node[right=6pt] at (5.05,1.50) {$\ell_{u,1}=1$};
		\node[right=6pt] at (7.05,0.50) {$\ell_{u,0}=2$};
		
	\end{tikzpicture}
	\caption{A Young diagram $Y_u$ is uniquely specified by the infinite sequence of nonnegative  integers $\{\ell_{u,0}, \ell_{u,1}, \ell_{u,2}, \ldots\}$ with $Y_{u,k}=0$ with sufficiently large $k$. For the diagram shown here, this sequence is $\{2,1,3,0,1,0,0,0,\ldots\}$, where the dots indicate that all subsequent entries are zero. The purple boxes have arm length zero and therefore form the set $Y_{u,0}$, while the gray boxes have arm length one and form the set denoted by $Y_{u,1}$.}
	\label{Y2}
\end{figure}
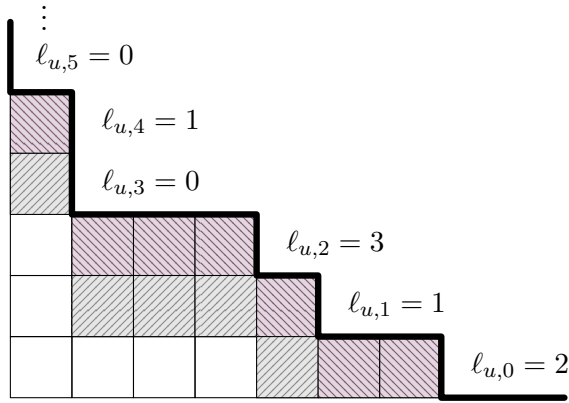

In this paper, we focus on the so-called light asymptotic limit, in which the central charge is sent to infinity by taking $b \to 0$, while the conformal dimensions remain finite. From (\ref{dim_gen}) it follows that this limit can be achieved by adopting the following parametrization of the conformal dimensions:
\bea
\label{etaDef}
\alpha_u = b\,\eta_u\,, \qquad
\lambda = b\,\mu\,,
\eea
with all parameters $\eta_u$ and $\mu$ kept finite.

Using (\ref{etaDef}), the AGT map (\ref{AGTmap}) can be rewritten as
\begin{eqnarray}
	\label{agt2}
	a_u = -\epsilon_1 \eta_u
	+ (\epsilon_1 + \epsilon_2)\left(\frac{n+1}{2} - u\right)\,,
	\qquad
	m = \epsilon_1 \frac{\mu}{n}\,.
\end{eqnarray}
In view of (\ref{bepsilon}), the small-$b$ limit is equivalent to the limit $\epsilon_1 \to 0$. We are therefore interested in the behaviour of (\ref{Fy}) in this regime. Indeed, substituting (\ref{agt2}) into (\ref{Zbf}) yields
\begin{eqnarray}
	\label{Zbf2}\nonumber
	Z_{bf}(a_u,Y_u\mid a_v,Y_v)=
	\displaystyle\prod_{s\in Y_u}\big(\epsilon_1(\eta_v-\eta_u+v-u-L_{Y_v}(s))+\epsilon_2(v-u+1+A_{Y_u}(s))\big)
	\\
	\times
	\displaystyle\prod_{s\in Y_v}\big(\epsilon_1(\eta_v-\eta_u+v-u+1+L_{Y_u}(s))+\epsilon_2(v-u-A_{Y_v}(s))\big)\,.
\end{eqnarray}
Taking this into account, it is easy to see that in the small-$\epsilon_1$ limit one obtains
\bea
\label{FsingleTerm}
\frac{Z_{bf}(a_u-m,Y_u\mid a_v,Y_v)}{Z_{bf}(a_u,Y_u\mid a_v,Y_v)}
&=&
\displaystyle\prod_{s\in Y_{u,u-v-1}}
\frac{\eta_v-\eta_u-\frac{\mu}{n}+v-u-L_{Y_v}(s)}
{\eta_v-\eta_u+v-u-L_{Y_v}(s)}
\\ \nonumber
&\times&
\displaystyle\prod_{s\in Y_{v,v-u}}
\frac{\eta_v-\eta_u-\frac{\mu}{n}+v-u+1+L_{Y_u}(s)}
{\eta_v-\eta_u+v-u+1+L_{Y_u}(s)}\,.
\eea

Here $Y_{u,r}$ denotes the set of boxes in the Young diagram $Y_u$ with arm length $r$ (see Fig.~\ref{Y2}). It is straightforward to verify that all other boxes do not contribute in this limit. We then derive the $W_n$ one-point conformal block by exploiting this formula. 

To avoid confusion, let us list all notation related to the  Young diagrams we have and will use throughout this paper.
\begin{itemize}
	\item $Y_{u,r}$ denotes the set of boxes in the Young diagram $Y_u$ with arm length $r$.
	\item $|Y_{u,r}|$ denotes the number of boxes with arm length $r$.
	\item 
	Any Young diagram $Y_u$ can be represented by a sequence of integers
	$\{\ell_{u,0},\ell_{u,1},\ldots\}$, where $\ell_{u,i}\in \{0,1,\ldots \}$, denotes a non‑negative integer. More precisely, we define
	\[
	\label{ellYrel}
	\ell_{u,k} := |Y_{u,k}| - |Y_{u,k+1}|,
	\]
	Notice that  $\ell_{u,k}=0$ when $k$ exceeds the number of rows in $Y_u$.
	For details see Fig.~\ref{Y2}.
\end{itemize}
\section{The Liouville one-point torus conformal block}
\label{Lblock}

For $n=2$, substituting (\ref{FsingleTerm}) into (\ref{Fy}) yields
\bea
\nonumber
F_{\{Y_1,Y_2\}}=
\displaystyle\prod_{s\in Y_{1,0}} \frac{-\frac{\mu}{2}+1+L_{Y_1}(s)}{1+L_{Y_1}(s)}
\displaystyle\prod_{s\in Y_{2,1}} \frac{\eta_2-\eta_1-\frac{\mu}{2}+2+L_{Y_1}(s)}{\eta_2-\eta_1+2+L_{Y_1}(s)}
\\ \label{FsingleTermLio}
\times
\displaystyle\prod_{s\in Y_{2,0}}
\frac{-\frac{\mu}{2}+1+L_{Y_2}(s)}{1+L_{Y_2}(s)}
\frac{\eta_1-\eta_2-\frac{\mu}{2}-1-L_{Y_1}(s)}{\eta_1-\eta_2-1-L_{Y_1}(s)}.
\eea
Using the notation introduced at the end of Section \ref{Zlight}, the factors in (\ref{FsingleTermLio}) can be rewritten as
\bea
\displaystyle\prod_{s\in Y_{1,0}} \frac{-\frac{\mu}{2}+1+L_{Y_1}(s)}{1+L_{Y_1}(s)}
=
\prod_{i=0}^{\infty} \frac{\left(-\frac{\mu}{2}+1\right)_{\ell_{1,i}}}{\ell_{1,i}!},
\eea
and similarly,
\bea
\displaystyle\prod_{s\in Y_{2,0}}
\frac{-\frac{\mu}{2}+1+L_{Y_2}(s)}{1+L_{Y_2}(s)}
=
\prod_{i=0}^{\infty} \frac{\left(-\frac{\mu}{2}+1\right)_{\ell_{2,i}}}{\ell_{2,i}!}.
\eea
Here $(x)_i$ denotes the Pochhammer symbol, defined as
\[
(x)_i = x(x+1)\cdots(x+i-1)\,.
\]
These formulas become transparent upon inspection of Fig.~\ref{Y2}. The remaining factors in (\ref{FsingleTermLio}) can be treated in the same way, although the result is somewhat less straightforward. One finds
\bea
\displaystyle\prod_{s\in Y_{2,0}}
\frac{\eta_1-\eta_2-\frac{\mu}{2}-1-L_{Y_1}(s)}
{\eta_1-\eta_2-1-L_{Y_1}(s)}
=
\prod_{i=0}^{\infty}
\frac{\left(\eta_2-\eta_1+\frac{\mu}{2}+1+S_{1,i}-S_{2,i}\right)_{\ell_{2,i}}}
{\left(\eta_2-\eta_1+1+S_{1,i}-S_{2,i}\right)_{\ell_{2,i}}},
\eea
and
\bea
\displaystyle\prod_{s\in Y_{2,1}}
\frac{\eta_2-\eta_1-\frac{\mu}{2}+2+L_{Y_1}(s)}
{\eta_2-\eta_1+2+L_{Y_1}(s)}
=
\prod_{i=0}^{\infty}
\frac{\left(\eta_2-\eta_1-\frac{\mu}{2}+2+S_{1,i}-S_{2,i+1}\right)_{\ell_{2,i+1}}}
{\left(\eta_2-\eta_1+2+S_{1,i}-S_{2,i+1}\right)_{\ell_{2,i+1}}}.
\eea

Here we introduce the notation
\bea
S_{u,k}=\sum_{i=k}^{\infty}\ell_{u,i}.
\eea
It is clear from (\ref{ellYrel}) and Fig.~\ref{Y2} that $S_{u,k}=|Y_{u,k}|$.

Substituting these expressions into (\ref{FsingleTermLio}), we obtain
\bea
\nn
F_{\{Y_1,Y_2\}}
=
\prod_{i=0}^{\infty}
\frac{\left(1-\frac{\mu}{2}\right)_{\ell_{1,i}}}{\ell_{1,i}!}
\frac{\left(1-\frac{\mu}{2}\right)_{\ell_{2,i}}}{\ell_{2,i}!}
\frac{\left(\eta_{2,1}+\frac{\mu}{2}+1+S_{1,i}-S_{2,i}\right)_{\ell_{2,i}}}
{\left(\eta_{2,1}+1+S_{1,i}-S_{2,i}\right)_{\ell_{2,i}}}
\\
\label{FY1Y2}
\times
\frac{\left(\eta_{2,1}-\frac{\mu}{2}+2+S_{1,i}-S_{2,i+1}\right)_{\ell_{2,i+1}}}
{\left(\eta_{2,1}+2+S_{1,i}-S_{2,i+1}\right)_{\ell_{2,i+1}}}.
\eea
where we have introduced the notation $\eta_{i,j}=\eta_i-\eta_j$.
As can be seen from Fig.~\ref{Y2} we also have
\bea
\label{qYtoql}
q^{|Y_1|+|Y_2|}
=
q^{\sum_{i=0}^{\infty} (i+1)\,(\ell_{1,i}+\ell_{2,i})}.
\eea
Therefore, substituting (\ref{FY1Y2}) and (\ref{qYtoql}) into (\ref{instfu}), we arrive at
\bea
\label{Zliou}
Z_{inst}
=
\sum_{\substack{\ell_{1,0},\ldots=0 \\ \ell_{2,0},\ldots=0}}^{\infty}
F_{\{Y_1,Y_2\}}\, q^{|Y_1|+|Y_2|}.
\eea
Notice that, at any fixed order in $q$, the sums in this expression and in~(\ref{qYtoql}), as well as the product in~(\ref{FY1Y2}), are all finite.

In \cite{Hadasz:2009db,Alkalaev:2016fok} it was shown  that\footnote{Full agreement is obtained after rescaling both $\mu$ and $\eta_{12}$ by a factor of two. This normalization follows from the AGT map of \cite{Wyllard:2009hg} and is required only in the Liouville case.}
\bea
\label{ZliouHyp}
Z_{inst}
=
(1-q)^{\frac{\mu}{2}}
\left(\prod_{n=1}^{\infty}(1-q^n)\right)^{\mu-2}
\,{}_2F_1\left(\frac{\mu}{2},\eta_{1,2}+\frac{\mu}{2}-1;\eta_{1,2};q\right).
\eea
This expression is considerably more elegant than our representation (\ref{Zliou}) together with (\ref{FY1Y2}) and (\ref{qYtoql}). In particular, (\ref{ZliouHyp}) has only simple poles at $\eta_{1,2}=-k$, where $k$ is any positive integer, whereas (\ref{FY1Y2}) also contains higher-order poles, as well as poles at negative values of $k$. We have checked that these two representations agree up to order twenty in $q$ (see the  Mathematica file attached).

A few remarks are in order. To show directly from (\ref{FY1Y2}) that the higher-order poles, as well as all poles with negative 
		$k$, cancel is not a simple task.
	The cancellation of   negative $k$ poles is expected  from the Zamolodchikov recursion relation for the torus conformal block, derived in \cite{Poghossian:2009mk,Hadasz:2009db}. More precisely, as noted in \cite{Poghossian:2009mk}, even for generic $b$, the poles of the individual terms in the instanton partition function, given by (\ref{instfu}) and (\ref{Fy}), are highly redundant: most of them disappear after summing all contributions at a fixed instanton order leaving only those associated with intermediate degenerate states (in 2d CFT terms). In this sense, the Zamolodchikov recursion relation exhibits a much simpler pole structure. On the contrary,  cancellation of higher-order poles with positive $k$ is not equally transparent, even from the recursion relation; to see it, one must first derive the recursion relation in the light limit.
	Notice also that  in \cite{Alba:2010qc}, the AGT conjecture was proven by regrouping and analyzing the contributions of appropriate sets of  Young diagrams. Thus, it is reasonable to expect that the pole cancellations in our case too  can be proven  along similar lines.

We also emphasize that, although the representation of the one-point light block in (\ref{ZliouHyp}) is much simpler than the one given by (\ref{Zliou}) together with (\ref{FY1Y2}), this simplification may be peculiar to the case $n=2$. Indeed, only for $n=2$ can the one-point torus block be mapped to the four-point block on the sphere. It is well known that, in the light limit, instanton counting for the four-point block on the sphere leads directly to a hypergeometric function \cite{Mironov:2009qn,Poghosyan:2016lya}.

Furthermore, our representation has the important advantage that it extends to $W_n$ with $n>2$ without substantial additional effort.

Before turning to this generalization, let us perform a simple consistency check. It is easy to see that, in the limit $\eta_{12}\to\infty$, both (\ref{ZliouHyp}) and (\ref{Zliou}) reduce to
\bea
Z_{inst}\, \substack{\approx \\ \eta_1 \to \infty}
\left(\prod_{n=1}^{\infty}(1-q^n)\right)^{\mu-2}.
\eea

\section{$W_n$ one-point light conformal  block}
\label{Wblock}

As in Liouville theory, the two factors appearing in Eq.~(\ref{FsingleTerm}) can be rewritten as
\bea
\nn
&& \prod_{s\in Y_{u,u-v-1}}
\frac{\eta_{v,u}-\frac{\mu}{n}+v-u-L_{Y_v}(s)}
{\eta_{v,u}+v-u-L_{Y_v}(s)}
=
\\
&&\qquad \prod_{i=0}^{\infty}
\frac{\left(\eta_{v,u}-\frac{\mu}{n}+v-u-S_{v,i}+S_{u,i+u-v-1}\right)_{\ell_{u,i+u-v-1}}}
{\left(\eta_{v,u}+v-u-S_{v,i}+S_{u,i+u-v-1}\right)_{\ell_{u,i+u-v-1}}}.
\eea
and
\bea
\nn
&& \prod_{s\in Y_{v,v-u}}
\frac{\eta_{v,u}-\frac{\mu}{n}+v-u+1+L_{Y_u}(s)}
{\eta_{v,u}+v-u+1+L_{Y_u}(s)}
=
\\
&& \qquad \prod_{i=0}^{\infty}
\frac{\left(\eta_{v,u}-\frac{\mu}{n}+v-u+1+S_{u,i}-S_{v,i+v-u}\right)_{\ell_{v,i+v-u}}}
{\left(\eta_{v,u}+v-u+1+S_{u,i}-S_{v,i+v-u}\right)_{\ell_{v,i+v-u}}}.
\eea

Substituting these expressions into Eq.~(\ref{Fy}), we obtain
\begin{align}
	\label{FyLight}
	F_{\vec{Y}}
	&=
	\prod_{i=0}^{\infty} \prod_{v=1}^{n}\prod_{u=1}^{v}
	\frac{\left(\eta_{v,u}-\frac{\mu}{n}+v-u+1+S_{u,i}-S_{v,i+v-u}\right)_{\ell_{v,i+v-u}}}
	{\left(\eta_{v,u}+v-u+1+S_{u,i}-S_{v,i+v-u}\right)_{\ell_{v,i+v-u}}}
	\nonumber\\
	&\quad\times
	\prod_{u=2}^{n}\prod_{v=1}^{u-1}
	\frac{\left(\eta_{v,u}-\frac{\mu}{n}+v-u+S_{u,i+u-v-1}-S_{v,i}\right)_{\ell_{u,i+u-v-1}}}
	{\left(\eta_{v,u}+v-u+S_{u,i+u-v-1}-S_{v,i}\right)_{\ell_{u,i+u-v-1}}}.
\end{align}

This expression can be rewritten in the equivalent form
\begin{align}
	\label{FyLight_fin}
	F_{\vec{Y}}
	=
	\prod_{i=0}^{\infty} \prod_{v=1}^{n}
	\frac{\left(1-\frac{\mu}{n}\right)_{\ell_{v,i}}}{\ell_{v,i}!}
	\prod_{u=1}^{v-1}
	\frac{\left(\eta_{v,u}-\frac{\mu}{n}+v-u+1+S_{u,i}-S_{v,i+v-u}\right)_{\ell_{v,i+v-u}}}
	{\left(\eta_{v,u}+v-u+1+S_{u,i}-S_{v,i+v-u}\right)_{\ell_{v,i+v-u}}}
	\nonumber\\
	\quad\times
	\frac{\left(\eta_{v,u}+\frac{\mu}{n}+v-u+S_{u,i}-S_{v,i+v-u-1}\right)_{\ell_{v,i+v-u-1}}}
	{\left(\eta_{v,u}+v-u+S_{u,i}-S_{v,i+v-u-1}\right)_{\ell_{v,i+v-u-1}}}.
\end{align}

Similarly to the Liouville case (\ref{qYtoql}), we have
\begin{align}
	\label{qYtoql_n}
	q^{|\vec{Y}|}
	=
	q^{\sum_{i=0}^{\infty} (i+1)\,(\ell_{1,i}+\ell_{2,i}+\dots+\ell_{n,i})}.
\end{align}
Therefore, by inserting the last two expressions into (\ref{instfu}), we obtain the 
instanton partition function:
\begin{align}
	\label{Zinst_n}
	Z_{\text{inst}}
	=
	\sum_{\substack{\ell_{1,0},\ldots=0\\ \dots \\ \ell_{n,0},\ldots=0}}^{\infty}
	F_{\vec{Y}}\, q^{|\vec{Y}|}.
\end{align}
Another representation for the case $n=3$ was obtained in \cite{Belavin:2024nnw}; for further details, see Appendix~\ref{W3block}.

In the previous section, we discussed the case $n=2$ and presented several arguments showing why this case is special and why our formula takes a more complicated form than the hypergeometric representation. We also argued that, for larger values of $n$, our representation may in fact be the most efficient one.

To support this claim, let us recall that, in general, the coefficients of the conformal block become highly intricate when one expands it as a series in $q$. It is only after multiplication by the $U(1)$ factor that this structure acquires the compact factorized form found in \cite{Alday:2009aq,Wyllard:2009hg}.

Let us now consider the next simplest case, namely $n=3$. The representation (\ref{Fshadow}) together with (\ref{fsh}) has both advantages and disadvantages compared to our formula. Its main advantage is its simpler pole structure. Its disadvantage, however, is that at each order in $q$ the coefficients involve rather complicated factors. As $n$ increases, one would naturally expect these structures to become even more involved. By contrast, this is not the case for our formula, whose form remains comparatively elegant for generic $n$.
\section{Summary}
\label{summary}

Since the natural first step toward a complete quantum formulation is the study of the quasiclassical limit, conformal blocks in this regime have been extensively investigated over the past decades. For Liouville theory and several of its generalizations, see \cite{Braaten:1983np,Braaten:1982yn,Thorn:2002am,Fateev:2007ab,Seiberg:1990eb,Zamolodchikov:1995aa,Fateev:2010za,Menotti:2006gc,Mironov:2009qn,Fateev:2011qa,Hama:2013ama,Poghosyan:2015oua,Poghosyan:2017qdv,Apresyan:2017few,Belavin:2018hfm,Poghosyan:2020zzg,Fioravanti:2019vxi,Fioravanti:2019awr,Cipriani:2025ikx}. The derivation of the one-point torus conformal block belongs to this line of work. The $W_2$ and $W_3$ cases considered here were previously obtained in \cite{Alkalaev:2016fok} and \cite{Belavin:2024nnw}, respectively. In the present paper,  we studied the one-point torus conformal block in $A_{n-1}$ Toda field theory in the light asymptotic limit, that is, in the large-central-charge regime with finite conformal dimensions. Using the AGT correspondence, we reformulated the problem in terms of the instanton partition function of four-dimensional ${\cal N}=2^\ast$ supersymmetric gauge theory with gauge group $U(n)$.

Our main observation is that, in the light limit, the instanton partition function simplifies substantially: only particular subsets of boxes in the Young diagrams contribute to the instanton sum. This makes it possible to derive an explicit representation for the $W_n$ one-point light conformal block, valid for arbitrary $n\ge 2$. For $n=2,3$, this representation has a more intricate pole structure than the previously known expressions. Nevertheless, it is far more efficient than first performing instanton counting and only then taking the light asymptotic limit. In fact, as $n$ increases, instanton counting becomes progressively more difficult because of its growing technical complexity. Our formula therefore makes it possible to obtain substantially higher-order instanton contributions (for $n=2,3,4,5$, one can reach twenty instantons without difficulty). We have attached a Mathematica file that can be used to generate these results.

We have also argued that, as $n$ increases, our representation is likely to be of the simplest form one can expect. It therefore provides a useful framework for investigating the large-$n$ behavior of one-point light conformal blocks on the torus.

\acknowledgments

The research of A.P. was supported by the Armenian SCS grant 21AG-1C060. Similarly, H.P.'s work was supported by the Armenian SCS grants 21AG-1C062
and 24WS-1C031.

\begin{appendix}
	\section{Another representation for the $W_3$ light conformal block}
	\label{W3block}
	
	In \cite{Belavin:2024nnw}, a representation for the $W_3$ one-point light block on the torus was obtained using the so-called shadow formalism. The result reads
	\bea
	\label{Fshadow}
	\,{\cal F}_{\bm{\eta}}^{\mu}(q)= (1-q)^{\frac{\mu }{3}-2} \frac{N_{gl}}{\tilde{N}}\sum _{i,k,l,m=0}^{\infty } \sum _{j=0}^i \sum _{n=0}^k  (-1)^{i+k+n} q^{i+j+k+l+m}f \, ,
	\eea
	where
	\bea
	\label{fsh}
	&f=
	C_j^i C_n^k C_i^{2-2 \eta _1-\eta _2-\frac{\mu }{3}} C_l^{\eta _1+2 \eta _2+\frac{\mu }{3}-2} C_m^{-\eta _1-2 \eta _2+\frac{\mu }{3}}
	\Gamma \left(j+n+\frac{\mu }{3}\right)\qquad \;
	\\ \nn
	& \times
	\frac{ \Gamma \left(k+\frac{\mu }{3}\right) \Gamma \left(1-k-\frac{\mu }{3}\right) \Gamma \left(k-l+m+n+\frac{\mu }{3}+\eta _1+2 \eta _2-1\right)\Gamma \left(j-k+l-m+\eta _1-\eta _2-\frac{\mu }{3}\right)}{\Gamma (k+1) \Gamma \left(i+2 \eta _1+\eta _2-1\right) \Gamma \left(j+n+2 \eta _1+\eta _2-1\right) \Gamma \left(-l+m+n+\eta _1+2 \eta _2\right)}
	\\ \nn
	& \times \;
	\Gamma \left(i-j+k-l+m+\frac{\mu }{3}+\eta _1+2 \eta _2-1\right)
	\,{}_3F_2\!\left(\begin{matrix}A,B,C\\ B_1,B_2\end{matrix};1\right) .
	\eea
	We also use the following notation:
	\begin{footnotesize}
		\bea
		\label{ABC}
		A=\eta _1-\eta _2+j-k+l-\frac{\mu }{3}-m,
		\\ \nn
		B=-\eta _1-2 \eta _2-k+l-\frac{\mu }{3}-m+2,
		\quad
		C=-\eta _1-2 \eta _2+l-m-n+1,
		\\ \nn
		B_1=-\eta _1-2 \eta _2-i+j-k+l-\frac{\mu }{3}-m+2,
		\quad
		B_2=-\eta _1-2 \eta _2-k+l-\frac{\mu }{3}-m-n+2 \, .
		\eea
	\end{footnotesize}
	Furthermore,
	\bea
	\nn
	\tilde{N}&=&\frac{\Gamma \left(1-\frac{\mu }{3}\right) \Gamma \left(\frac{\mu }{3}\right)^2 \Gamma \left(-\frac{\mu }{3}+\eta _1-\eta _2\right) \Gamma \left(\frac{\mu }{3}+\eta _1+2 \eta _2-1\right){}^2}{\Gamma \left(2 \eta _1+\eta _2-1\right){}^2 \Gamma \left(\eta _1+2 \eta _2\right)}
	\\
	\label{Ntilde}
	&\times & \, _2F_1\left(-\eta _1-2 \eta _2+1,-\frac{\mu }{3}+\eta _1-\eta _2;-\frac{\mu }{3}-\eta _1-2 \eta _2+2;1\right),
	\eea
	and
	\bea
	N_{gl}=(1-q)^3 (q+1)\left(q^{-\frac{1}{24}}\eta(q)\right){}^{-2}.
	\eea
	The binomial coefficient is denoted by
	\bea
	C^i_j=\binom{i}{j}.
	\eea
	We have compared this representation of the conformal block with our result up to third order in $q$. In principle, obtaining higher-order terms in both representations is straightforward. The main difficulty arises in comparing them. Indeed, the expression in (\ref{Fshadow}) contains the hypergeometric function $_3F_2$ evaluated at the  point $1$, which makes the analysis difficult even at the numerical level. Details of the comparison are provided in the attached Mathematica file.

	Let us make a few remarks concerning the use of formulas (\ref{Fshadow}), (\ref{fsh}), and (\ref{Ntilde}). There are two hypergeometric functions that require special care. For the first one, we may use the well-known Gauss summation formula to obtain
	\bea
	& \, _2F_1\left(1-\eta _1-2 \eta _2,\eta _1-\eta _2-\frac{\mu }{3};2-\frac{\mu }{3}-\eta _1-2 \eta _2;1\right)
	=\frac{
		\Gamma\!\left(2-\frac{\mu}{3}-\eta_1-2\eta_2\right)\,
		\Gamma\!\left(1-\eta_1+\eta_2\right)
	}{
		\Gamma\!\left(1-\frac{\mu}{3}\right)\,
		\Gamma\!\left(2-2\eta_1-\eta_2\right)
	},
	\eea
	Let us now analyze the hypergeometric function $_3F_2$. From (\ref{ABC}) one sees that $B_1$ and $B_2$ differ from $B$ only by integer shifts:
	\bea
	B_1=B-i+j\,,
	\qquad
	B_2=B-n \, .
	\eea
	The expression can be simplified by applying Thomae's transformation (see for example \cite{rathie2016newprooffundamentaltwoterm}),
	\bea
	{}_3F_2\!\left(\begin{matrix}a,b,c\\ b_1,b_2\end{matrix};1\right)
	=
	\frac{\Gamma(b_1)\Gamma(b_2)\Gamma(s)}
	{\Gamma(a)\Gamma(s+b)\Gamma(s+c)}
	\,{}_3F_2\!\left(\begin{matrix} b_1-a,\ b_2-a,\ s\\ s+b,\ s+c\end{matrix};1\right),
	\eea
	where
	\[
	s=b_1+b_2-a-b-c.
	\]
	This gives
	\bea
	\label{hypId}
	&{}_3F_2\!\left(\begin{matrix}B,A,C\\ B-i+j,B-n\end{matrix};1\right)
	=
	\frac{\Gamma(B-i+j)\Gamma(B-n)\Gamma(S)}
	{\Gamma(B)\Gamma(S+A)\Gamma(S+C)}
	\,{}_3F_2\!\left(\begin{matrix}j-i,\ -n,\ S\\ S+A,\ S+C\end{matrix};1\right),
	\eea
	where
	\[
	S=\eta _2-\eta _1-i-l+m+1.
	\]

	Now, inspecting (\ref{Fshadow}), we see that both $j-i$ and $-n$ are non-positive integers. Therefore, the hypergeometric function on the right-hand side of (\ref{hypId}) truncates to a finite sum. This is the form that we used when working with the one-point light block on the torus in the representation (\ref{Fshadow}). In the attached Mathematica file, we compared this result with ours and found them to be in agreement.
\end{appendix}

\bibliographystyle{JHEP}
\providecommand{\href}[2]{#2}\begingroup\raggedright\endgroup

\end{document}